\documentclass[12pt]{article}
\usepackage{amssymb}
\usepackage{verbatim}

\newcommand{\X}{\ensuremath{{\cal{X}} \, }}

\newcommand{\N}{\ensuremath{{\cal{N}} \, }}

\newcommand{\C}{\ensuremath{{\cal{C}} \, }}

\newcommand{\M}{\ensuremath{{\cal{M}} \, }}

\newcommand{\Real}{\ensuremath{{\mathbb{R}} \, }}

\newcommand{\cL}{\ensuremath{{\cal{L}} \, }}

\newcommand{\p}{\partial}

\newcommand{\ie}{\emph{i.e.},}
\newcommand{\eg}{\emph{eg.},}

\begin{document}

\begin{titlepage}

\title{\bf{Classical History Theory of Vector Fields}}
\author{Duncan Noltingk\footnote{e-mail d.noltingk@ic.ac.uk}\\
        \\
        Blackett Laboratory\\
        Imperial College\\
        Prince Consort Road\\
        London SW7 2BZ}


\maketitle

\begin{abstract}
 We consider the extension of classical history theory to the
 massive vector field and electromagnetism. It is argued that
 the action of the two Poincare groups introduced by Savvidou
 suggests that the history fields should have
 five components. The extra degrees of freedom introduced to make
 the fields five-dimensional result in an extra pair of second
 class constraints in the case of the massive vector field, and in
 an extended gauge group in the case of electromagnetism. The
 total gauge transformations depend on two arbitrary parameters,
 and contain `internal' and `external' $U(1)$ gauge transformations as
 subgroups.
\end{abstract}

\end{titlepage}

\section{Introduction}

\subsection{Motivation}

 The Hamiltonian formalism provides a strong link between
 classical and quantum theories and is mathematically
 well-developed in both cases. However, a major drawback of the
 standard Hamiltonian approach to relativistic field theories
 is that it requires a splitting of space-time into space and
 time, thus breaking the manifest covariance of the theory. This
 problem becomes particularly acute when canonical methods are
 applied to generally covariant theories such as general relativity,
  and this is one aspect of the
 `problem of time' in canonical quantum gravity.

\smallskip
 Recently a Hamiltonian formalism has been proposed by Isham and
 co-workers \cite{Ish1,Ish2,Ish3},
 in which the fundamental physical entities are
 entire histories of the system under consideration, as opposed to Cauchy data at an
 instant of time. For a thorough introduction see \cite{sav1}.
 The histories formalism was originally developed
 in the quantum case, but there is a well-defined
 classical history formalism. The central object in a
 classical history theory is the space of histories, $\Pi$, which
 is defined to be a one-parameter family of single-time phase
 spaces. An element of $\Pi$ is called a history, and the space of
 histories carries a symplectic structure which provides the crucial
 link to the corresponding quantum history theory.

\smallskip
 A particularly
 striking fact about history theories is that there are two
 notions of time evolution\cite{sav2}. External time corresponds to the
 parameter labeling the copies of state space, and internal time
 is generated by Hamiltonian evolution. In this way the notion of
 time plays two different roles in a history theory, one
 corresponding to the causal ordering of logical propositions, and
 the second corresponding to dynamical evolution. The two times
 are linked together in the action principle.

\smallskip
 The histories formalism has been applied to scalar field theory
 on flat \cite{sav3} and curved \cite{anas,nolt} space-times, and
 in the following we discuss
 the extension of the classical theory to the case of vector fields.
 In particular we examine the
 massive vector field and the electromagnetic field from a
 histories perspective. We shall argue that, as a consequence of
 the two time directions, a history field
 should be regarded as a certain type of vector field on a
 five-dimensional space-time. We also give an analysis of the
 gauge transformations of histories electromagnetism.

\smallskip
 These results are relevant to the ultimate goal of the
 histories programme: the formulation of a histories version of
 general relativity \cite{sav4}. Firstly, they suggest the
 possibility that a covariant history theory of gravity should be
 concerned with the metric of a five-dimensional extended
 space-time. Secondly, gauge symmetry is an important feature of
 general relativity, particularly when formulated in terms of
 tetrad fields. The study of the extended gauge transformations of
 histories electromagnetism is relevant in this context.

\subsection{Scalar Field Theory}

 In this section we give a brief account of the histories
 description of the classical scalar field on Minkowski
 space-time, \M, as given by Savvidou \cite{sav3}. We define the
 Minkowski metric $\eta_{\mu\nu}$ to have signature $(+,-,-,-)$.

\subsubsection{The history algebra}
 The canonical configuration space of the scalar field is $Q_{n,t} =
 C^\infty(\Sigma_{n,t})$ where $\Sigma_{n,t}$ is a Cauchy surface
 in \M. The Cauchy surfaces are labeled by a future pointing timelike unit
 vector $n$ normal to $\Sigma_{n,t}$, and a real number $t$. Each
 Cauchy surface represents an instant of time in a particular
 inertial frame. The state space $P_{n,t}$ is the cotangent bundle
 of $Q_{n,t}$ which we identify with $C^\infty(\Sigma_{n,t}) \times
 C^\infty(\Sigma_{n,t})$\footnote{The analytical subtleties
 regarding these infinite dimensional spaces will not concern us
 here. We shall only be interested in the resulting Poisson
 algebra.}.
 The construction of the corresponding
 history theory begins by defining a trivial vector bundle
 \footnote{In this paper we will not make use of this bundle
 structure, and we could just have defined $\Pi_n$ to be the space
 of paths $\Real \rightarrow P_n$. However, the bundle picture is
 useful in more general situations (\eg \, on curved space-times),
 and also gives a motivation for the `internal' / `external'
 nomenclature in history theory: internal transformations act
 internally to the fibres of $\xi_n$, while external transformations
 act across the fibres.}
  $\xi_n : P \times
 \Real \rightarrow \Real$ such that $\xi_n^{-1}(t) = P_{n,t}$.
 Here $P$ is an abstract copy of the state space.
 The space of histories of the scalar field corresponds to the space of sections
 of this bundle, $\Pi_n = \Gamma(\xi_n)$, and the history fields
 satisfy the following Poisson algebra:
 \begin{eqnarray}
 \{ \phi_n(t;\underline{x}) , \phi_n(t';\underline{x'}) \} &=& 0 \\
 \{ \pi_n(t;\underline{x}) , \pi_n(t';\underline{x'}) \} &=& 0 \\
 \{ \phi_n(t;\underline{x}) , \pi_n(t';\underline{x'}) \} &=&
 \delta(t-t') \delta_{n,t}^{(3)}(\underline{x}-\underline{x'})
 \end{eqnarray}
 where $\delta_{n,t}^{(3)}(\underline{x}-\underline{x'})$ is the delta
 function on $\Sigma_{n,t}$. The right hand side of this equation
 is a space-time scalar density of weight one. In Minkowski space-time, there is
 no difference between a scalar density and a scalar (if we
 restrict attention to transformations under the connected part
 of the Poincare group), which suggests
 that the fields can be thought of as space-time scalars.
 A pair $(t,\underline{x}) \in \Real \times \Sigma$ can be
 identified with a unique four-vector $X \in \M$ as $X=tn+x_n$,
 where the three-vector $\underline{x}$ has been associated with a
 corresponding four-vector $x_n$ that is $n$-spatial (\ie \, $n \cdot x_n
 := \eta_{\mu\nu}n^\mu x_n^\nu = 0$). Hence we can
 write $\phi_n(t;\underline{x})$ as $\phi_n(X)$. The history
 algebra can then be written in the more covariant looking form
 \begin{eqnarray}
 \label{covalg}
 \{ \phi_n(X) , \phi_n(X') \} &=& 0 \\
 \{ \pi_n(X) , \pi_n(X') \} &=& 0 \\
 \{ \phi_n(X) , \pi_n(X') \} &=& \delta^{(4)}(X-X')
 \end{eqnarray}
 Now it is tempting to drop the $n$ label from the fields since the
 right hand side of this algebra does not depend
 on $n$. However, this would be somewhat misleading in the
 sense that the field $\pi(X)$ has no physical meaning. This is
 because the conjugate momentum corresponds to the field momentum along a
 certain timelike \emph{direction}, and so must be
 written as $\pi_n(X)$. Although the algebra (\ref{covalg}), (5), (6) is
 independent of $n$, Savvidou \cite{sav3} has shown that in the
 quantum theory $n$ labels the physically relevant, inequivalent
 \emph{representations} of the algebra on a particular Fock space.
 So in the analysis of the classical theory the
 $n$-labels remain on the fields on the understanding that they
 are necessary for the physical interpretation of the theory
 and arise naturally in quantisation. However, we note that it is rather
 unsatisfactory to have an $n$ label on the $\phi$ field because,
 physically, the value of the field at a point in space-time
 \emph{is} independent of the foliation. In this histories formulation of
 classical
 scalar field theory all propositions about the field are made
 in the context of a particular inertial reference frame\footnote{The
 `multisymplectic' approach to field theory offers a way round
 this problem as it concerns a scalar field $\phi$ and a
 \emph{vector} field $\pi_\mu$ \cite{Ish4}.}.

\subsubsection{Time translations}
 For each $n$, a `Louville' operator
 can be constructed from the fields,
 \begin{equation}
 V_n := \int d^4X \pi_n n_\mu \p^\mu \phi_n
 \end{equation}
 which generates external time translations in the $n$ direction.
 In coordinates adapted to $n$ these transformations take the
 form $\phi_n(t,\underline{x}) \mapsto \phi_n(t+\lambda,\underline{x})$,
  and similarly for $\pi_n(t,\underline{x})$.

\smallskip
 There is another notion of time translation in the history
 theory. Intuitively the Hamiltonian at each instant of external
 time generates dynamical evolution \emph{internal} to each fibre of
 $\xi_n$. More precisely, the time-averaged Hamiltonian
 \begin{equation}
 H_n = \frac{1}{2} \int d^4X [\pi_n^2 + (\eta^{\mu\nu}-n^\mu n^\nu)\p_\mu
 \phi_n \p_\nu \phi_n + m^2 \phi_n^2]
 \end{equation}
 generates transformations $\phi_n(X) \mapsto \phi_n(X,s)$.

\smallskip
 The action operator is made up of the Louville and Hamiltonian
 operators as follows:
 \begin{equation}
 S_n := V_n - H_n
 \end{equation}
 and the equations of motion can be written in the form
 \begin{eqnarray}
 \{ S_n , \phi_n(X) \} = 0 \\
 \{ S_n , \pi_n(X) \} = 0
 \end{eqnarray}

\subsubsection{Poincare covariance}
 Savvidou \cite{sav3} has shown the existence of two Poincare groups in the
 histories formulation of the scalar field. The $n$-spatial
 components of the two
 groups are identical, but the time translations of the internal
 Poincare group are generated by the Hamiltonian while the
 time translations of the external Poincare group are
 generated by the Louville operator.

\smallskip
 External boosts correspond to the following automorphism of the
 history algebra:
 \begin{eqnarray}
 \phi_n(t,\underline{x},0) \mapsto \phi_{\Lambda n}(\Lambda
 (t,\underline{x}),0) \\
 \pi_n(t,\underline{x},0) \mapsto \pi_{\Lambda n}(\Lambda
 (t,\underline{x}),0)
 \end{eqnarray}
 where we have used adapted coordinates to write $\phi_n(X,0) =
 \phi_n(t,\underline{x},0)$ and $\Lambda(t,\underline{x})$ denotes the usual Lorentz
 transformations acting on inertial coordinates $(t,\underline{x})$. We note
 that, in the classical case, these automorphisms cannot be generated by
 canonical transformations. This is because there is no
 momentum conjugate to the foliation vector $n$, and thus no
 way to generate changes in $n$. The `multisymplectic' approach
 \cite{Ish4} offers a solution to this problem in the classical
 theory, and in the quantum theory changes in $n$ correspond to
 mapping between inequivalent representations of the history
 algebra.

\smallskip
 Internal boosts act on the fields as follows:
 \begin{eqnarray}
 \phi_n(0,\underline{x},s) \mapsto
 \phi_n(0,\Lambda(\underline{x},s)) \\
  \pi_n(0,\underline{x},s) \mapsto
  \pi_n(0,\Lambda(\underline{x},s))
 \end{eqnarray}
 As the internal boosts leave the foliation vector fixed they can
 be implemented by canonical transformations.
 The generator of internal boosts on the hyperplane $s=const$ is
 \begin{equation}
 {}^{int}K_n(m) = m_\mu \int d^4X [\pi s \p^\mu \phi - X^\mu
 H_n(X)]
 \end{equation}
 where $H_n(X)$ is the Hamiltonian density and the integral is
 over the surface $s=const$.

 \smallskip
 The fields $\phi_n(X,s)$ are defined on an extended space-time $\N
 = \M \times \Real$. However, the theory is not invariant under the full
 $SO(2,3)$ isometry group of this space-time. This is evident from
 the fact that the algebra (\ref{covalg}), (5), (6) is defined on external space-time,
 that is the surface in \N defined by $s=0$, and not on internal
 space-time (defined by $n \cdot X=0$). Also, the generators of the
 symmetries are defined as integrals over external space-time. This
 indicates that the fields are not true scalar fields on \N.
 However, the fields are invariant under the internal and
 external $SO(1,3)$ subgroups of $SO(2,3)$.

 In the case of the scalar field these subtleties can be
 overlooked, but the construction of a history theory of
 vector fields
 acutely illustrates this issue. A particularly relevant
 question is whether the history vector field should have
 $4$ indices, or $5$ as it must to be a vector field
 on \N.

\section{Massive vector field}

\subsection{State space theory}

 In this section we give a brief overview of the standard state
 space theory of the massive vector field on \M.

\smallskip
 We begin with the covariant theory. The covariant configuration
 space is $\X(\M)$, the space of vector fields on \M.
 The massive vector field is described by the following
 Lagrangian \cite{chang}:
 \begin{equation}
 \cL = -\frac{1}{4} \phi_{\mu\nu} \phi^{\mu\nu} + \frac{1}{2} m^2
 \phi_\mu \phi^\mu
 \end{equation}
 where $\phi^{\mu\nu}(X) := \p^{[\,\mu} \phi^{\nu]}(X)$ and $\phi
 \in \X(\M)$.
 The resulting field equations are
 \begin{eqnarray}
 (\square + m^2) \phi^\mu(X) &=& 0 \\
 \p_\mu \phi^\mu(X) &=& 0.
 \end{eqnarray}
 The first of these equations shows that each component of the
 field behaves like a massive scalar field. The second equation is
 known as the Fierz-Pauli equation and it is the first indication
 of the presence of constraints in the theory.

\smallskip
 To pass to the canonical theory we choose a Cauchy surface in \M
 and consider the fields on this Cauchy surface. More precisely, we
 choose a space-like embedding $\iota : \Sigma \rightarrow \M$, where
 $\Sigma \simeq \Real^3$, and take the corresponding configuration space
 to be the space of fields $\phi^\mu_\iota(\underline{x})$ where
 $\underline{x} \in \iota(\Sigma)$. However the fields
 $\phi_\iota^\mu(\underline{x})$ are not geometric objects on either
 $\Sigma$ or \M. The geometrical interpretation of the fields is
 clarified by considering $\mbox{Emb}(\Sigma,\M)$, the space
 of embeddings of $\Sigma$ into \M.
 Then $\phi_\iota$ can be thought of as an element of  $T_\iota
 \mbox{Emb}(\Sigma,\M)$ where the tangent space to
 $\mbox{Emb}(\Sigma,\M)$ at the embedding $\iota$ is defined as
 \begin{equation}
 T_\iota \mbox{Emb}(\Sigma,\M) = \{ \psi : \Sigma \rightarrow T\M
 | \psi(\underline{x}) \in T_{\iota(\underline{x})}\M\}
 \end{equation}
 The configuration space of the canonical theory is then defined
 as $Q_\iota := T_\iota \mbox{Emb}(\Sigma,\M)$ for some fixed $\iota$.
 The cotangent space of $\mbox{Emb}(\Sigma,\M)$ at $\iota$
 is defined similarly:
 \begin{equation}
 T^*_\iota \mbox{Emb}(\Sigma,\M) = \{ l : \Sigma \rightarrow T^*\M
 | l(\underline{x}) \in T^*_{\iota(\underline{x})}\M\}
 \end{equation}
 and the pairing between these two spaces is given by
 \begin{equation}
 <l , \psi >_\iota = \int_\Sigma d\theta_{\underline{x}} \, l_\mu (\iota(\underline{x})) \psi^\mu
 (\iota(\underline{x}))
 \end{equation}
 where $d\theta_{\underline{x}}$ is an arbitrary volume element on
 $\Sigma$.
 The state space $P_\iota$ is the cotangent bundle of $Q_\iota$, and can be
 identified with the Cartesian product $Q_\iota \times T^*_\iota
 \mbox{Emb}(\Sigma,\M)$.

\smallskip
 As we are considering flat space-time,
 there exists a family of preferred embeddings; those which
 correspond to inertial frames. The space of preferred embeddings
 can be parametrised by pairs $(n,t)$ where $n$ is a  future
 pointing unit vector in \M and $t \in \Real$. As we are
 considering a deterministic system, we choose $t=0$ without loss.
 We denote the configuration space and the state space corresponding
 to the embedding labeled by $(n,0)$ as $Q_n$ and $P_n$
 respectively. The state space $P_n$ carries the following Poisson
 algebra;
 \begin{eqnarray}
 \{ \phi_n^\mu(\underline{x}) , \phi_n^\nu(\underline{x}') \} &=& 0 \\
 \{ \pi^n_\mu(\underline{x}) , \pi^n_\nu(\underline{x}') \} &=& 0 \\
 \{ \phi_n^\mu(\underline{x}) , \pi^n_\nu(\underline{x}') \} &=&
 \delta^\mu_\nu \delta_n^{(3)}(\underline{x}-\underline{x}')
 \end{eqnarray}
 where $\delta_n^{(3)}(\underline{x}-\underline{x}')$ is the delta
 function on $\Sigma_n$.

\smallskip
 A field $\phi_n^\mu(\underline{x}) \in Q_n$ can be decomposed into
 the pair
 $(\phi_n^t(\underline{x}) , {}^n \phi^\mu(\underline{x}))$ where
 \begin{eqnarray}
 \phi_n^t(\underline{x}) &:=& n_\mu \phi^\mu_n(\underline{x}) \\
 {}^n\phi^\mu(\underline{x}) &:=& {}^nP^\mu_\nu \phi^\nu_n(\underline{x})
 \end{eqnarray}
 and we have introduced the $n$-spatial projection tensor
 defined by
 \begin{equation}
 {}^n P^\mu_\nu := \delta^\mu_\nu - n^\mu n_\nu
 \end{equation}
 It follows that $n_\mu \, {}^nP^\mu_\nu = 0$ and $n^\nu \,
 {}^nP^\mu_\nu = 0$. The fields $\phi_n^t(\underline{x})$ and
 ${}^n\phi^\mu(\underline{x})$ are defined on the space of
 embeddings, but $\phi_n^t(\underline{x})$ pulls back to give
 a scalar field on $\Sigma_n$. We can use the metric on \M to
 lower the index on ${}^n\phi^\mu(\underline{x})$. The resulting
 one-form can be pulled back to $\Sigma_n$, and then the index can
 be raised using the metric on $\Sigma_n$ to give a vector field
 on $\Sigma_n$\footnote{Although this is an inherently non-linear
 process, it presents no extra difficulties in the case of Minkowski space-time.}.
 In a similar way, objects defined by
 ${}^n\pi_\mu(\underline{x}) := {}^nP_\mu^\nu
 \pi^n_\nu(\underline{x})$, and
 ${}^n\phi^{\mu\nu}(\underline{x}) := {}^nP^\mu_\sigma \,
 {}^nP^\nu_\rho \phi_n^{\sigma\rho}(\underline{x})$ can be thought
 of as a one-form and a covariant tensor on $\Sigma_n$.

\smallskip
 The canonical momenta are computed from the Lagrangian and
 turn out to be \cite{tyutin}
 \begin{eqnarray}
 \pi^n_t(\underline{x}) = 0 \,\,\, ,
 \,\,\, {}^n\pi_\mu(\underline{x}) =
 {}^nP^\rho_\mu n^\nu \phi_{\nu\rho}(\underline{x})
 \end{eqnarray}
 The first of these equations is a primary constraint.
 The canonical Hamiltonian is computed to be
 \begin{eqnarray}
 H_n = \int_{\Sigma_n} d\theta_{\underline{x}} \, [ \frac{1}{2}{}^n\pi_\mu \, {}^n\pi^\mu
 - \frac{1}{4} {}^n\phi_{\mu\nu}\, {}^n\phi^{\mu\nu}
 - \frac{1}{2}m^2 (\phi_n^t)^2  \\ +  \frac{1}{2}m^2 \, {}^n\phi_\mu \, {}^n\phi^\mu -
 \phi_n^t \, {}^n\partial^\mu \, {}^n\pi_\mu ]
 \end{eqnarray}
 where the $n$-spatial derivative is defined as ${}^n \p^\mu = {}^n P^\mu_\nu
 \p^\nu$.
 For the primary constraint $\pi^n_t(\underline{x}) = 0$ to be preserved by the
 dynamical evolution, it is necessary and sufficient that
 $\{H_n,\pi^n_t(\underline{x})\}=0$. This implies the secondary constraint
 \begin{equation}
 m^2 \phi_n^t(\underline{x}) + {}^n\partial^\mu \, {}^n\pi_\mu(\underline{x}) = 0.
 \end{equation}
 A point in the constraint surface determines and is fully determined by
 the pair $({}^n\pi_\mu(\underline{x}) , {}^n\phi^\nu(\underline{x}))$. The
 constraint surface is therefore isomorphic to the space
 $C_n = \Lambda^1(\Sigma_n) \times \X(\Sigma_n)$
 where $\Lambda^1(\Sigma_n)$ is the space of $1$-forms on
 $\Sigma_n$.

 The constraints do not commute under the Poisson bracket and
 so they form a second class pair. This implies that the pull-back
 of the symplectic $2$-form on $P_n$ to the constraint
 surface is non-degenerate. Therefore there is a well-defined
 Poisson algebra on the constraint surface which is given by the
 Dirac brackets:
 \begin{equation}
 \{{}^n\phi^\mu(\underline{x}) , {}^n\pi_\nu(\underline{x}')\}_D =
 {}^nP^\mu_\nu \delta^{(3)}(\underline{x}-\underline{x}')
 \end{equation}

 \subsection{Classical History theory}

 We begin with the abstract state space $P$, and follow the usual
 procedure of taking a one-parameter family of copies of $P$. This
 results in a trivial vector bundle $ \xi_n : P \times \Real \rightarrow
 \Real$ where the fibre $\xi_n^{-1}(t) = P_{n,t}$ for each
 $t$. Sections of $\xi_n$ correspond to histories of the vector
 field with respect to the foliation labelled by $n$.
 Thus a history is a map
 \begin{equation}
 h_n : t \mapsto (\phi_n^\mu(t;\underline{x}) , \pi^n_\nu(t;\underline{x}))
 \end{equation}
 If we choose a volume element on $\Real$ then
 a symplectic structure is induced on the space of
 maps $\Real \rightarrow P$ because $P$ is a symplectic manifold.
 The symplectic structure defines the following algebra
 \begin{eqnarray}
 \label{mapalg}
 \{\phi^\mu_n(t;\underline{x}) , \phi_n^\nu(t';\underline{x}')\}
  &=& 0 \\
 \{\pi_\mu^n(t;\underline{x}) , \pi^n_\nu(t';\underline{x}')\}
  &=& 0 \\
 \{\phi^\mu_n(t;\underline{x}) , \pi^n_\nu(t';\underline{x}')\}
 &=& \, \delta^\mu_\nu \delta(t-t')\delta^{(3)}_n (\underline{x}-\underline{x}')
 \end{eqnarray}
 In the case of the scalar field a one-parameter family of functions
 on $\Sigma_n$:
 \begin{equation}
 t \mapsto \phi_n(t;\underline{x})
 \end{equation}
 was identified with a function $\phi_n(X)$ on $\M$. In a
 similar way, for a fixed foliation,
 a one-parameter family of elements of $Q_n$
 \begin{equation}
 t \mapsto \phi_n^\mu(t;\underline{x})
 \end{equation}
 is equivalent to a unique vector field $\phi_n^\mu(X) \in \X(\M)$.
 The one-parameter family $ t \mapsto \pi^n_\mu(t;\underline{x})$
 can be identified with a one-form $\pi^n_\mu(X) \in \Lambda^1(\M)$
 in the same way, and these fields satisfy the covariant looking
 algebra
 \begin{eqnarray}
 \{\phi^\mu_n(X) , \phi_n^\nu(X')\} &=& 0 \\
 \{\pi_\mu^n(X) , \pi^n_\nu(X')\} &=& 0 \\
  \{\phi_n^\mu(X) , \pi^n_\nu(X')\} &=& \delta^\mu_\nu \delta^{(4)} (X-X')
 \end{eqnarray}
 In this way, the history space, $\Pi_n$, can be identified with the
 space $\X(\M) \times \Lambda^1(\M)$. We can decompose $\phi_n^\mu(X)$
 into the pair $(\phi_n^t(X) , {}^n\phi^\mu(X))$ defined by
 \begin{eqnarray}
 \phi_n^t(X) &:=& n_\mu \phi_n^\mu(X) \\
 {}^n\phi^\mu(X) &:=& {}^nP^\mu_\nu \phi_n^\nu(X)
 \end{eqnarray}
 and we use these fields to define the generators of internal and
 external time translations.
 External time translations are generated by the `Louville'
 operator,
 \begin{equation}
 V_n := \int d^4X \, [ \pi^n_t n_\nu \p^\nu \phi_n^t + {}^n \pi_\mu n_\nu \p^\nu \,
 {}^n \phi^\mu]
 \end{equation}
 Internal time translations are generated by the time-averaged
 Hamiltonian
 \begin{eqnarray}
 H_n = \int d^4X [
 \frac{1}{2} {}^n\pi_\mu \, {}^n\pi^\mu +
 \frac{1}{4} {}^n\phi_{\mu\nu}\, {}^n\phi^{\mu\nu} -
 \frac{1}{2} m^2 (\phi_n^t)^2      \\
 + \frac{1}{2} m^2 \, {}^n\phi_\mu \, {}^n\phi^\mu -
 \phi_n^t \, {}^n\partial^\mu \, {}^n\pi_\mu
 ]
 \end{eqnarray}
 The internal time translations generated by $H_n$ take the form
 \begin{equation}
 \phi_n^\mu(X) \mapsto \phi_n^\mu(X,s)
 \end{equation}
 In the next subsection we will consider the geometric meaning
 of these curious objects which have four components, but depend
 on five space-time variables.

\smallskip

 In the remainder of this subsection we discuss the history constraint
 surface $\C_n \subset \Pi_n$. An arbitrary element of $\Pi_n$ will not
 be compatible with the constraints. The constraint
 submanifold, $\C_n$, contains
 all elements of $\Pi$ which satisfy the constraints:
 \begin{equation}
 \pi^n_t(X)=0 \,\,\, , \,\,\, m^2 \phi_n^t(X) + {}^n \p^\mu \, {}^n \pi_\mu(X) = 0
 \end{equation}
 Thus a point in $\C_n$ is equivalent to a pair
 $({}^n\phi^\mu(X) , {}^n\pi_\nu(X))$.
  The Poisson
 algebra induced on $\C_n$ by pulling back along the natural
 inclusion map $\C_n \hookrightarrow \Pi_n$ is given by
 \begin{eqnarray}
 \{{}^n \phi^\mu(X) , {}^n \phi^\nu(X')\} &=& 0 \\
 \{{}^n \pi_\mu(X) , {}^n \pi_\nu(X')\} &=& 0 \\
 \{{}^n \phi^\mu(X) , {}^n \pi_\nu(X')\} &=& {}^nP^\mu_\nu \delta^{(4)}(X-X')
 \end{eqnarray}
 and we note that $\C_n$ is diffeomorphic to the space of sections of
 the bundle $C_n \times \Real \rightarrow \Real$.

 \subsection{Poincare covariance}

 As discussed in the previous section for the case of the scalar field, the notion of
 two times leads naturally to the definition of two Poincare
 groups. The external Poincare group mixes the external time with
 the $n$-spatial variables $\underline{x}$
 and the internal Poincare group mixes internal time with $\underline{x}$.
 The crucial
 new feature of the generators for the vector field is the
 mixing of the time-like and space-like components of the field.

\subsubsection{External Poincare group}
  The generators of external space-time translations
  can be written in covariant looking form as
 \begin{equation}
 {}^{ext}P^\mu_n = \int d^4X \pi^n_\nu(X) \p^\mu \phi_n^\nu(X)
 \end{equation}
 Next we define
 \begin{equation}
 M^{\mu\nu}_n = \int d^4X \pi^n_\rho(X) (X^\mu \p^\nu -
 X^\nu \p^\mu) \phi_n^\rho(X) + \sigma_n^{\mu\nu}
 \end{equation}
  where the `spin tensor' is
 \begin{equation}
 \sigma^{\mu\nu}_n = \int d^4X \pi^n_\rho(X) (\delta^{\rho \mu} \delta_\sigma^\nu -
 \delta^{\rho \nu} \delta_\sigma^\mu ) \phi_n^\sigma(X)
 \end{equation}
 The generators of $n$-spatial rotations\footnote{\ie \, rotations
 that leave $n$ fixed if $n$ is considered as a space-time vector
 rather than as the foliation label.} can be parametrised by
 two vectors $m^1$ and $m^2$ satisfying $m^1 \cdot n = m^2 \cdot n
 = 0$ in the following way:
 \begin{equation}
 M_n(m^1,m^2) = m^1_\mu m^2_\nu \,  M^{\mu\nu}_n
 \end{equation}
 As in the case of the scalar field, the external boosts cannot be
 implemented by canonical transformations because of the change in
 foliation. The natural definition
 of the automorphisms generated by the action of the external
 boosts is
 \begin{eqnarray}
 \phi_n^\mu(t,\underline{x},0) \mapsto \Lambda^\mu_\nu
 \phi_{\Lambda n}^\nu(\Lambda(t,\underline{x}),0) \\
 \pi^n_\mu(t,\underline{x},0) \mapsto \Lambda_\mu^\nu
 \pi^{\Lambda n}_\nu(\Lambda(t,\underline{x}),0)
 \end{eqnarray}
 So the external boosts mix $t$ with $\underline{x}$, and the external time
 component $\phi_{\Lambda n}^t(t,\underline{x},0)$
 with the space-like components ${}^{\Lambda n}\phi^\mu(t,\underline{x},0)$.

\subsubsection{Internal Poincare group}
 The rotation and spatial translation generators of the internal
 Poincare group coincide with those for the external Poincare group.
 However, internal time translations are generated by the Hamiltonian, and
 therefore act internally as $\phi_n^\mu(X) \mapsto
 \phi_n^\mu(X,s)$.
 We tentatively define the generator of internal Lorentz
 transformations on the $s=const$ hyperplane in the `obvious' way:
 \begin{equation}
 {}^{int}K_n(m) :=  m_\mu \int d^4X [\pi^n_\nu s \p^\mu \phi_n^\nu
 - X^\mu H_n(X)] + n_\mu m_\nu
 \sigma^{\mu\nu}
 \end{equation}
 where $H_n(X)$ is the Hamiltonian density, $m$ is a vector
 satisfying $m \cdot n = 0$, and the integral is
 over the surface $s=const$. This functional generates the automorphisms
 \begin{eqnarray}
 \phi_n^\mu(0,\underline{x},s) \mapsto \Lambda^\mu_\nu
 \phi_n^\nu(0,\Lambda(\underline{x},s)) \\
 \pi^n_\mu(0,\underline{x},s) \mapsto \Lambda_\mu^\nu
 \pi^n_\nu(0,\Lambda(\underline{x},s))
 \end{eqnarray}
 ${}^{int}K_n(m)$ mixes $\emph{s}$ with $\underline{x}$, and
 $\phi^t_n(0,\underline{x},s)$ with ${}^n\phi_n^\mu(0,\underline{x},s)$. So in this
 transformation, the function $\phi^t_n(0,\underline{x},s)$ is associated with
 the \emph{internal} time direction, whereas it was
 associated with the \emph{external} time direction by the
 external boosts.

\subsection{Alternative interpretations of the history vector
field}

 The above discussion suggests that it is misleading to think of the
 history vector fields as a family of $4$-vectors on
 external space-time. We propose two alternative interpretations
 of the history vector fields.

 \subsubsection{Four-component fields}

 One way of thinking of the history fields is as a family of
 $4$-vectors, but with the temporal component in the
 $\partial_\tau := \partial_s + \partial_t$ direction;
 \begin{equation}
 \phi_n = \phi_n^\tau(X,s) \partial_\tau + \phi_n^i(X,s) \partial_i
 \end{equation}
 where $\phi_n^\tau(X,s) = \phi_n^t(X,s)$ and we have used
 coordinates adapted to $n$.
 From this perspective it is natural to look for a representation
 of the Poincare group in which the boosts act in this direction.
 The orbital part of the internal boost generator on $s=0$ would be
 \begin{equation}
 K_n(m) := \int d^4X [n \cdot X \pi^n_\mu(X) m_\nu \partial^\nu \phi_n^\mu(X)
  - m \cdot X (V_n(X) + H_n(X))]
 \end{equation}
 where $V_n(X)$ is the `Louville' density.
 However, it can be shown that
 \begin{equation}
 \{ K_n(m^1) , K_n(m^2) \} \neq m^1_\mu m^2_\nu M_n^{\mu\nu}
 \end{equation}
 even on the solutions to the equations of motion, and so the
 functionals $K_n(m)$ do not form a representation of the
 Poincare group.

\smallskip
 Thus it is not possible to eliminate the two times in
 favour of one `physical' time direction $\partial_\tau$,
 in a covariant way. Nevertheless, $\tau$ does have a
 special significance in the theory. This is indicated by the
 equations of motion. Returning to the example of the scalar field
 for a moment; if $\phi_n(X)$
 is a solution then $\{S_n , \phi_n(X)\} = 0$ implies that
 \begin{equation}
 (\partial_s - \partial_t)\phi_n(X)|_{s=0} = 0
 \end{equation}
 and so all the temporal change in such histories occurs
 in the $\tau$ direction.

 \subsubsection{Five-component fields}

 We can augment $\phi_n^\mu(X,s)$ with a new degree of freedom
 $\phi_n^s(X,s)$ to form
 $\tilde{\phi}_n^M(X,s)$, a five component `vector field' on the
 extended space-time $\N = \M \times \Real$ with metric
 $\mbox{diag}(+,-,-,-,+)$. The extended fields
 are written
 \begin{equation}
 \tilde{\phi}_n = \tilde{\phi}_n^M(X,s) \partial_M  \,\,\, , \,\,\,
 \tilde{\pi}^n = \tilde{\pi}^n_M(X,s) dx^M
 \end{equation}
 The label $M$ runs over $t,1,2,3,s$, where we define $M=t$ to refer
 to external time, $M=1,2,3$ to correspond to the spatial directions,
 and $M=s$ to refer to internal time.

 The extended
 history space, $\tilde{\Pi}_n \subset \X(\N) \times \Lambda^1(\N)$,
 contains all pairs $(\tilde{\phi}_n,\tilde{\pi}_n) \in \X(\N) \times
 \Lambda^1(\N)$ that satisfy the internal field equations:
 \begin{eqnarray}
 \p_s \tilde{\phi}_n^M(X,s) = \{ \tilde{H}_n , \tilde{\phi}_n^M(X,s) \} \\
 \p_s \tilde{\pi}^n_M(X,s) = \{ \tilde{H}_n , \tilde{\pi}^n_M(X,s) \}
 \end{eqnarray}
 where $\tilde{H}_n$ is the Hamiltonian on extended history space, and
 is defined in the next section.

 The history
 algebra can be extended to these fields in a natural way:
 \begin{equation}
 \{\tilde{\phi}_n^M(X) , \tilde{\pi}^n_N(X')\} = \delta^M_{\,N} \delta^{(4)}(X-X')
 \end{equation}
 This
 defines the algebra of the fields on the surface $s=0$, which is a
 submanifold of \N. Hamiltonian evolution can be used
 to extend this definition to the rest of \N. The fact that the
 algebra is naturally defined on the hyperplane $s=0$ and not on the
 hyperplane $n \cdot X=0$ reflects the underlying asymmetry between the
 two `modes' of time. As a consequence of this asymmetry, the
 theory constructed from the fields $\tilde{\phi}_n$ will not be covariant
 under the full $SO(2,3)$ isometry group associated with \N. In
 particular it will not be covariant under the action of the
 $SO(2)$ subgroup acting in the $(s,t)$ plane.

\smallskip
 In order to discuss the Poincare transformations of the five-component fields,
 we make the definition:
 \begin{equation}
 \tilde{M}_n^{MN} = \int_{\M^{(e)}} d^4X \tilde{\pi}^n_R(X) (X^M \p^N -
 X^N \p^M) \tilde{\phi}_n^R(X) + \tilde{\sigma}_n^{MN}
 \end{equation}
  where $\M^{(e)} \subset \N$ is external space-time, defined as
 the surface $s=0$, and the extended spin tensor is defined as
 \begin{equation}
 \tilde{\sigma}_n^{MN} = \int_{\M^{(e)}} d^4X \tilde{\pi}^n_A(X) (\delta^{A M} \delta_B^N -
 \delta^{A N} \delta_B^M ) \tilde{\phi}_n^B(X)
 \end{equation}
 Using $\tilde{M}_n^{MN}$ we can write the rotation generators as
 \begin{equation}
 \tilde{M}_n(\tilde{m}^1,\tilde{m}^2) = \tilde{m}^1_M \tilde{m}^2_N
 \tilde{M}_n^{MN}
 \end{equation}
 where $\tilde{m}^1$ and $\tilde{m}^2$ are $n$-spatial in the sense that
 $\tilde{m}^1_M \tilde{n}^M = \tilde{m}^1_M \tilde{e}^M = 0$, and
 similarly for $\tilde{m}^2$.

\smallskip
 The foliation vector $n$ is an element of the external
 space-time which is a subspace of \N. Using the canonical
 embedding, $n$ can be considered as a vector in \N, which we
 denote by $\tilde{n}$ and is given in coordinates $(X,s)$ by
 $(n , 0)$. Using this coordinate system the internal
 future pointing  unit vector $\tilde{e}$ can be written as $(0,1)$.
 We can use these vectors to decompose an extended field
 $\tilde{\phi}_n^M(X,s)$ into its external and internal time components as
 follows:
 \begin{equation}
 \tilde{\phi}_n^t(X,s) = \tilde{n}_M \tilde{\phi}_n^M(X,s) \,\,\, , \,\,\,
 \tilde{\phi}_n^s(X,s) = \tilde{e}_M \tilde{\phi}_n^M(X,s)
 \end{equation}
 Finally, given three orthogonal, $n$-spatial, unit vectors in \N,
 $\tilde{m}^i$, where
 $i=1,2,3$, the spatial components of
 $\tilde{\phi}_n^M(X,s)$ are
 \begin{equation}
 \tilde{\phi}_n^i(X,s) = \tilde{m}^i_M \tilde{\phi}_n^M(X,s)
 \end{equation}
 Using this basis we write the action of the external boosts
 as
 \begin{eqnarray}
 \tilde{\phi}_n^s(t,\underline{x},0) &\mapsto& \ \tilde{\phi}_{\Lambda n}^s(\Lambda(t,\underline{x}),0)  \\
 \tilde{\phi}_n^\mu(t,\underline{x},0) &\mapsto& \ \Lambda_\nu^\mu
 \tilde{\phi}_{\Lambda n}^\nu(\Lambda(t,\underline{x}),0) \\
 \tilde{\pi}^n_s(t,\underline{x},0) &\mapsto& \ \tilde{\pi}^{\Lambda n}_s(\Lambda(t,\underline{x}),0)  \\
 \tilde{\pi}^n_\mu(t,\underline{x},0) &\mapsto& \ \Lambda^\nu_\mu
\tilde{\pi}^{\Lambda n}_\nu(\Lambda(t,\underline{x}),0)
 \end{eqnarray}
 where $\mu$ takes the values $t,1,2,3$.

\smallskip
 Internal boosts are generated by
\begin{equation}
 {}^{int}\tilde{K}_n(\tilde{m}) := \tilde{m}_M \int_{\M_s^{(e)}} d^4X
 [ \tilde{\pi}^n_N s \, \p^M \tilde{\phi}_n^N - X^M \tilde{H}_n(X)] +
 \tilde{e}_M \tilde{m}_N \tilde{\sigma}_n^{MN}
 \end{equation}
and the resulting automorphisms are
 \begin{eqnarray}
 \tilde{\phi}_n^t(0,\underline{x},s) &\mapsto&
 \tilde{\phi}_n^t(0,\Lambda(\underline{x},s))  \\
 \tilde{\phi}_n^{\bar{\mu}}(0,\underline{x},s) &\mapsto&
  \Lambda_{\bar{\nu}}^{\bar{\mu}}
  \tilde{\phi}_n^{\bar{\nu}}(0,\Lambda(\underline{x},s))  \\
 \tilde{\pi}^n_t(0,\underline{x},s) &\mapsto&
 \tilde{\pi}^n_t(0,\Lambda(\underline{x},s))  \\
 \tilde{\pi}^n_{\bar{\mu}}(0,\underline{x},s) &\mapsto&
  \Lambda^{\bar{\nu}}_{\bar{\mu}}
  \tilde{\pi}^n_{\bar{\nu}}(0,\Lambda(\underline{x},s))
 \end{eqnarray}
 where $\bar{\mu}$ takes the values $1,2,3,s$.
  Now the components of $\tilde{\phi}_n$ and $\tilde{\pi}^n$
  are mixed in a way which is
 consistent with the mixing of the space-time variables. This
 indicates that the extended fields are an appropriate way of
 thinking about the history fields. However, it should be emphasised that the
 extended fields are not covariant under $SO(2,3)$.
 Let $V^M$ denote the vector defined by
 the field $\tilde{\phi}_n^M(X,s)$
 at the point $(X,s)$. Using the basis $(\tilde{n},\tilde{e},\tilde{m}^i)$,
  $V^M$ can be decomposed into
 an `external' four-vector $(V^t , V^i)$, or into an
 `internal' four-vector $(V^s , V^i)$, and each
 of these four-vectors \emph{is} a covariant object under the appropriate
 Poincare group. This suggests a third interpretation of the
 history vector fields as \emph{pairs} of
 four-component fields with identical
 $n$-spatial components. However, this identification is not
 preserved under the action of the external boosts, so it seems
 that we are left with the five-vector interpretation as the only
 viable one.

 It remains to be shown that
 the extra degrees of freedom can be included in the action in
 a way that is consistent with the symmetries and equations of motion of the
 theory.

\subsection{The action}

 In the case of scalar field theory the physical action functional
 is written as $S_n=V_n-H_n$. The Louville operator is associated with
 external time in the sense that it generates translations in the
 external time direction. In the same way, the Hamiltonian is
 associated with internal time, and the action functional mixes the
 two `modes' of time.

\smallskip
 First we will need the following definition:
 The $n$-spatial part of $\tilde{\phi}_n^M(X,s)$ is defined as
 \begin{equation}
 {}^n \tilde{\phi}^M(X,s) = {}^n \tilde{P}^M_N \tilde{\phi}^N(X,s)
 \end{equation}
 where the extended $n$-spatial projection tensor is
 \begin{equation}
 {}^n \tilde{P}^M_N := \delta^M_N - \tilde{n}^M \tilde{n}_N -
 \tilde{e}^M \tilde{e}_N
 \end{equation}
 Similarly we define ${}^n \p^M := {}^n\tilde{P}^M_N \p^N$.

 Let $\tilde{V}_n$ denote the extension of the Louville operator to the
 extended fields. $\tilde{V}_n$ is defined in the following natural way,
 \begin{equation}
 \tilde{V}_n := \int_{\M^{(e)}} d^4X [\tilde{\pi}^n_s \p_n^t
 \tilde{\phi}_n^s +
 \tilde{\pi}^n_t \p_n^t \tilde{\phi}_n^t + {}^n\tilde{\pi}_M \p_n^t
 {}^n\tilde{\phi}^M]
 \end{equation}
 where $\p_n^t = \tilde{n}_M \p^M$ is the derivative in the
 external time direction defined by $\tilde{n}$.
 The extended Hamiltonian, $\tilde{H}_n$, is defined as
 \begin{eqnarray}
 \tilde{H_n} := \int_{\M^{(e)}} d^4X [ \frac{1}{2}{}^n\tilde{\pi}_M \, {}^n\tilde{\pi}^M +
 \frac{1}{4} {}^n\tilde{\phi}_{MN}\, {}^n\tilde{\phi}^{MN} -
 \frac{1}{2} m^2 (\tilde{\phi}_n^s)^2  \\
 - \frac{1}{2} m^2 (\tilde{\phi}_n^t)^2 +
 \frac{1}{2} m^2 \, {}^n\tilde{\phi}_M \, {}^n\tilde{\phi}^M -
 \tilde{\phi}_n^t \, {}^n\partial^M \, {}^n\tilde{\pi}_M]
 \end{eqnarray}
 The important thing about this Hamiltonian is that
 $\tilde{\phi}_n^s$ and  $\tilde{\phi}_n^t$ both appear in the mass
 term, but only $\tilde{\phi}_n^t$ appears as the coefficient of
 ${}^n\partial^M \, {}^n\tilde{\pi}_M$. Due to this asymmetry
 between $\tilde{\phi}_n^s$ and $\tilde{\phi}_n^t$, the
 Hamiltonian is not invariant under $SO(2,3)$.
 The action is defined to be $\tilde{S}_n := \tilde{V}_n -
 \tilde{H}_n$. The resulting field equations are,
 \begin{eqnarray}
 \{\tilde{S}_n,\tilde{\phi}_n^s\} = 0 \,\,\, &\Rightarrow& \,\,\,
 \p_n^t \tilde{\phi}_n^s = 0 \\
 \{\tilde{S}_n,\tilde{\pi}^n_s\} = 0 \,\,\, &\Rightarrow& \,\,\,
 \p_n^t \tilde{\pi}^n_s + m^2 \tilde{\phi}_n^s = 0 \\
 \{\tilde{S}_n,\tilde{\phi}_n^t\} = 0 \,\,\, &\Rightarrow& \,\,\,
 \p_n^t \tilde{\phi}^t = 0 \\
 \{\tilde{S}_n,\tilde{\pi}^n_t\} = 0 \,\,\, &\Rightarrow& \,\,\,
 \p_n^t \tilde{\pi}^n_t + m^2 \tilde{\phi}_n^t + {}^n\p^M \, {}^n\tilde{\pi}_M = 0 \\
 \{\tilde{S}_n,{}^n\tilde{\phi}^M\} = 0 \,\,\, &\Rightarrow& \,\,\,
 \p_n^t \, {}^n \tilde{\phi}^M - ({}^n\tilde{\pi}^M + {}^n\p^M \tilde{\phi}_n^t) = 0 \\
 \{\tilde{S}_n,{}^n\tilde{\pi}_M\} = 0 \,\,\, &\Rightarrow& \,\,\,
 \p_n^t \, {}^n \tilde{\pi}_M - ({}^n\p^N \, {}^n \tilde{\phi}_{NM} + m^2 \, {}^n\tilde{\phi}_M) = 0
 \end{eqnarray}
 The physical action has not been derived in the usual way from a
 Lagrangian, so we do not have the usual identification of primary
 constraints. The field equations do not determine the
 time-like components of the $\tilde{\pi}$ field, so we augment the
 equations of motion with the following equations which are
 interpreted as the primary constraints of the theory.
 \begin{equation}
 \tilde{\pi}^n_s(X) = 0 \,\,\, , \,\,\, \tilde{\pi}^n_t(X) = 0
 \end{equation}
 We require these constraints to be conserved in internal time
 which implies the following secondary constraints
 \begin{equation}
 \tilde{\phi}_n^s(X) = 0  \,\,\, , \,\,\, m^2\tilde{\phi}_n^t(X)
 + {}^n\p^M \, {}^n\tilde{\pi}_M(X) = 0
 \end{equation}
 so in the history theory of the massive vector field there are
 \emph{two} pairs of second class constraints. In the state space
 theory a single pair of constraints allow the theory to be
 written in a Lorentz covariant way. In the history theory where
 there are two $SO(1,3)$ symmetry groups, we have to introduce two pairs of
 constraints in order to have a covariant theory.

 \bigskip

 The three functionals, $\tilde{V}_n$, $\tilde{H}_n$ and $\tilde{S}_n$ are all
 invariant under the action of the internal Poincare
 group. However, the external boosts change the foliation with
 respect to which $\tilde{V}_n$ and  $\tilde{H}_n$ are defined, giving
 the transformations $\tilde{V}_n \mapsto \tilde{V}_{\Lambda n}$
 and $\tilde{H}_n \mapsto \tilde{H}_{\Lambda n}$.
 These transformations imply $\tilde{S}_n \mapsto \tilde{S}_{\Lambda
 n}$, and the history theory is covariant under both
 Poincare groups if we include the internal foliation dependence.

\section{Electromagnetism}
\subsection{State space theory}
 In this section we consider vacuum electromagnetism on $\M$.
 The covariant configuration space for electromagnetism is $Q = \Lambda^1 (\M)$,
 and the Lagrangian is
 \begin{equation}
 \cL :=  - \frac{1}{4} F_{\mu\nu}F^{\mu\nu}
 \end{equation}
 where $F_{\mu\nu}(X) := \p_{[\mu} A_{\nu]}(X)$ for $A \in Q$.
 The covariant equations of motion which follows from this
 Lagrangian are
 \begin{equation}
 F^{\mu\nu}_{\,\,\,\,\,\, , \, \nu}(X) = 0
 \end{equation}
 Given an embedding $\iota: \Sigma \rightarrow \M$, the canonical
 configuration space is $Q_\iota = T^*_\iota \mbox{Emb}(\Sigma,\M)$ and the
 state space, $P_\iota$, can be identified with
 $Q_\iota \times T_\iota \mbox{Emb}(\Sigma,\M)$.
 The Poisson algebra on the state space associated with the embedding
 labeled by $(n,0)$ is
 \begin{eqnarray}
 \{A^n_\mu(\underline{x}) , A^n_\nu(\underline{x}')\} &=& 0 \\
 \{E_n^\mu(\underline{x}) , E_n^\nu(\underline{x}')\} &=& 0 \\
 \{A^n_\mu(\underline{x}) , E_n^\nu(\underline{x}')\} &=&
 \delta_\mu^\nu \delta_n^{(3)}(\underline{x}-\underline{x}')
 \end{eqnarray}
 The canonical momenta are computed from the Lagrangian;
 \begin{equation}
 E^t_n(\underline{x}) = 0 \,\,\, , \,\,\,
 {}^nE^\mu(\underline{x}) = {}^nP^\mu_\rho n_\nu
 F^{\nu\rho}(\underline{x})
 \end{equation}
 where we again use the decomposition into time-like and
 space-like parts. The canonical Hamiltonian is
 \begin{equation}
 H_n = \int_{\Sigma_n} d\theta_{\underline{x}} \, [\frac{1}{2}{}^nE^\mu \, {}^nE_\mu
 - \frac{1}{4} {}^nF_{\mu\nu} \, {}^nF^{\mu\nu} - A^n_t
 {}^n\p_\mu {}^n\pi^\mu]
 \end{equation}
 The equation of motion for $E^t_n$ implies the secondary constraint
 \begin{equation}
 {}^n\p_\mu \, {}^nE^\mu(\underline{x}) = 0
 \end{equation}
 which can be recognised as Gauss' law. The constraints form a
 first class pair. The corresponding
 gauge freedom is manifested in the fact that the equations of
 motion do not determine $A^n_t$ or $A^n_L$ where $A^n_L$ is the
 longitudinal part of $A^n$. The pull back of the symplectic
 two-form on $P_n$ to the constraint surface $C_n$ is degenerate
 because of the first class nature of the constraints. The reduced
 state space is obtained after gauge fixing and contains only
 four of the original eight degrees of freedom.

\subsection{History theory}
 We follow the same procedure as before and consider
 sections of the trivial vector bundle $P \times \Real
 \rightarrow \Real$. So a history is a map
 \begin{equation}
 t \mapsto (A_\mu^n(t;x) , E^\nu_n(t;x))
 \end{equation}
 and again, there is a unique pair $(A^n_\mu(X) , E^\nu_n(X))$
 corresponding to such a history so we fix a $n$ and identify
 $\Pi_n$ with the abstract space $\Lambda^1(\M) \times \X(\M)$.
 These history fields satisfy the covariant looking algebra,
 \begin{eqnarray}
 \label{algm}
 \{ A^n_\mu (X) , A^n_\nu (X') \} &=& 0 \\
 \{ E_n^\mu (X) , E_n^\nu (X') \} &=& 0 \\
 \{ A^n_\mu (X) , E_n^\nu (X') \} &=&
 \delta^\mu_\nu \, \delta^{(4)}(X - X')
 \end{eqnarray}
 and the Louville, Hamiltonian and action operators are defined as
 \begin{eqnarray}
 V_n &:=& \int d^4X [E_n^t \p_n^t A^n_t + {}^nE^\mu \p_n^t \, {}^nA_\mu] \\
  H_n &:=& \int d^4X [{}^nE_\mu \, {}^nE^\mu +
    {}^nF_{\mu\nu} \, {}^nF^{\mu\nu} -
    A^n_t \, {}^n\partial^\mu \, {}^nE_\mu] \\
  S_n &:=& V_n - H_n
 \end{eqnarray}

 At this point we note that,
 unlike in the case of the scalar field, the foliation dependence
 of these fields is empirically verified. It is a well-known fact
 that observations of magnetic and electric fields do depend on
 the state of motion of the observer.

\subsection{Constraints}

 We follow the procedure detailed for the vector field and
 work in the extended history space $\tilde{\Pi}_n \subset \Lambda^1
 (\N) \times \X(\N)$. The extended history fields are written in the form
 \begin{equation}
 \tilde{A}^n = \tilde{A}^n_M(X,s) dx^M  \,\,\, , \,\,\,
 \tilde{E}_n = \tilde{E}_n^M(X,s) \p_M
 \end{equation}
 and using a basis we have the decomposition of $\tilde{A}^n$
 into time-like and space-like components $(\tilde{A}^n_s , \tilde{A}^n_t ,
 \tilde{A}^n_i)$.
  The action of the Poincare group is
 very similar to the corresponding definitions for the vector
 field. From now on we fix an $n$, and work in coordinates
 adapted to $n$, dropping the $n$-label for typographical
 convenience.
 The new feature of electromagnetism is, of course, gauge
 invariance. We extend the Louville, Hamiltonian and action
 functionals to the extended history space as
 \begin{eqnarray}
 \tilde{V} &:=& \int_{\M^{(e)}} d^4X [
 \tilde{E}^s \p_t \tilde{A}_s +
 \tilde{E}^t \p_t \tilde{A}_t +
 \tilde{E}^i \p_t \tilde{A}_i
 ]  \\
 \tilde{H} &:=& \int_{\M^{(e)}} d^4X [
  \frac{1}{2} \tilde{E}_i \tilde{E}^i +  \frac{1}{4} \tilde{F}_{ij} \,
  \tilde{F}^{ij} - \tilde{A}_t  \partial^i \tilde{E}_i
 ] \\
  \tilde{S} &:=& \tilde{V} - \tilde{H}
 \end{eqnarray}
 As before we regard the equations
 \begin{equation}
 \label{constraint1}
 \tilde{E}^s(X) = 0 \,\,\, , \,\,\, \tilde{E}^t(X) = 0
 \end{equation}
 as the primary constraints of the theory. The corresponding
 secondary constraints follow from the Hamilitionian evolution of
 $\tilde{E}^s$ and $\tilde{E}^t$. The equation $\{\tilde{H},\tilde{E}^s(X)\} = 0$ is
 identically satisfied and $\{\tilde{H},\tilde{E}^t(X)\} = 0$ implies Gauss'
 law,
 \begin{equation}
 \label{constraint2}
 \p_i \tilde{E}^i(X) = 0
 \end{equation}
 Gauss' law is conserved in internal time, $\{\tilde{H},\p_i \tilde{E}^i(X)\} =
 0$, as a consequence of the anti-symmetry of $\tilde{F}_{ij}$.
 The equations (\ref{constraint1}) and (\ref{constraint2})
 are the first class constraints of the theory.

 \subsection{External local symmetries}

 To investigate the external local symmetries we
 define the extended action \cite{henn}.
 \begin{equation}
 \tilde{S}^E := \tilde{S} - \int_{\M^{(e)}} d^4X [\lambda_0 \tilde{E}^s +
 \lambda_1 \tilde{E}^t + \lambda_2 \p_i \tilde{E}^i]
 \end{equation}
 The transformations
 \begin{eqnarray}
 \delta \tilde{A}_s(X) &=& \epsilon_0(X) \\
 \delta \tilde{A}_t(X) &=& \epsilon_1(X) \\
 \delta \, \tilde{A}_i(X) &=& \p_i \epsilon_2(X)
 \end{eqnarray}
 are generated by the functional
\begin{equation}
 \psi = \int_{\M^{(e)}} d^4X [\epsilon_0 \tilde{E}^s +
 \epsilon_1 \tilde{E}^t + \epsilon_2 \p_i \tilde{E}^i]
 \end{equation}
 The extended action is invariant under these transformations if
 \begin{eqnarray}
 \delta \lambda_0(X) &=& \p^t\epsilon_0(X) \\
 \delta \lambda_1(X) &=& \p^t\epsilon_1(X) \\
 \delta \lambda_2(X) &=& \epsilon_1(X) - \p^t\epsilon_2(X)
 \end{eqnarray}
 The symmetry of the total action, and therefore of the
 underlying Lagrangian theory is found by setting $\lambda_2 = 0$
 (and $\delta \lambda_2 = 0$),
 thus eliminating the secondary constraint. The resulting
 transformations are
 \begin{eqnarray}
 \label{exgauge1}
 \delta \tilde{A}_s(X) &=& \epsilon_0(X) \\
 \label{exgauge2}
 \delta \tilde{A}_\mu(X) &=& \p_\mu \epsilon_2(X)
 \end{eqnarray}
 where $\mu = t,1,2,3$. These transformations contain two
 arbitrary real functions on external space-time.
 Setting $\epsilon_0 = 0$ we obtain `external' $U(1)$ gauge
 transformations:
 \begin{equation}
 \delta \tilde{A}_\mu(X) = \p_\mu \epsilon_2(X)
 \end{equation}
 which correspond to the symmetries of the external Maxwell tensor
 $\tilde{F}_{\mu \nu} = \p_{[ \, \mu}
 \tilde{A}_{\nu \, ]}$.

  \subsection{Internal local symmetries}
 The map $\tilde{A}_M(X) \mapsto \tilde{A}_M(X,s)$ is one-to-many for gauge
 systems. This introduces an extra ambiguity into the theory which
 is not contained in equations (\ref{exgauge1}) and (\ref{exgauge2}).
 We make this extra ambiguity
 explicit by introducing Lagrange multipliers to make the map
 $\tilde{A}_M(X) \mapsto \tilde{A}_M(X,s)$ one-to-one. To accomplish this we
 define the extended Hamiltonian;
 \begin{equation}
 \tilde{H}_s^E[\lambda_0^s,\lambda_1^s,\lambda_2^s] :=
 \tilde{H} - \int_{\M_s^{(e)}} d^4 X [\lambda^s_0 \tilde{E}^s +
 \lambda_1^s \tilde{E}^t  + \lambda^s_2 \p_i \tilde{E}^i]
 \end{equation}
 At each moment of internal time $s$, the integral is over $\M_s^{(e)}$ (the
 surface $s=\mbox{const}$), and the Lagrange multipliers are arbitrary real
 valued functions $\lambda^s : \M_s^{(e)} \rightarrow \Real$.
 The extended Hamiltonian generates canonical transformations of
 the potential field
 \begin{equation}
 \label{extendeom}
 \p_s \tilde{A}_M(X,s) = \{ H^E_s , \tilde{A}_M(X,s)\}
 \end{equation}
 and the map $\tilde{A}_M(X) \mapsto \tilde{A}_M(X,s)$ is given by the
 flow of the time-dependent vector field generated by $H^E_s$.
 Thus $\tilde{A}_M(X,s)$ is the solution of the following integral
 equation
 \begin{equation}
 \tilde{A}_M(X,s) = \tilde{A}_M(X,0) +
 \mbox{exp}\left( \int_0^s ds' \{\tilde{H}^E_{s'} , \tilde{A}_M(X,s')\} \right)
 \end{equation}
 Because the transformation $\tilde{A}_M(X) \mapsto
 \tilde{A}_M(X,s)$ is canonical, it preserves the Poisson
 bracket so
\begin{equation}
 \{ \tilde{A}_M(X,s) , \tilde{E}^N(X',s)\} = \delta_M^N
 \delta^{(4)} (X - X')
\end{equation}
 Gauge-equivalent histories correspond to different choices of the
 Lagrange multipliers in the extended Hamiltonian. The functional
 defined by
 \begin{equation}
 \psi_s = \int_{\M_s^{(e)}} d^4X [\epsilon^s_0 \tilde{E}^s +
 \epsilon^s_1 \tilde{E}^t + \epsilon^s_2 \p_i \, \tilde{E}^i]
 \end{equation}
 generate transformations on $\M_s^{(e)}$ as follows:
 \begin{equation}
 \delta \tilde{A}_M(X,s) = \{\psi_s , \tilde{A}_M(X,s)\}
 \end{equation}
 and these transformations take the form
 \begin{eqnarray}
 \delta \tilde{A}_s(X,s) &=& \epsilon^s_0(X) \\
 \delta \tilde{A}_t(X,s) &=& \epsilon^s_1(X) \\
 \delta \tilde{A}_i(X,s) &=& \p_i \epsilon^s_2(X)
 \end{eqnarray}
 In order that equation (\ref{extendeom}) is preserved by these
 transformations up to a change in the Lagrange multipliers
 associated with the primary constraints,
 the transformations must satisfy
 \begin{eqnarray}
 \delta \tilde{A}_s(X,s) &=& \epsilon^s_0(X) \\
 \delta \tilde{A}_t(X,s) &=& \p_s \epsilon^s_2(X) \\
 \delta \tilde{A}_i(X,s) &=& \p_i \epsilon^s_2(X)
 \end{eqnarray}
 These are the internal local symmetry transformations of histories
 electromagnetism. They contain two arbitrary real valued functions
 on extended space-time, but are not just a
 trivial extension of the external local transformations to each
 moment of internal time. The transformation of $\tilde{A}_t$
 contains a derivative with respect to internal time rather than
 external time. This is because the internal gauge transformations
 correspond to the symmetries of the internal field equations
 rather than the symmetries of the external field equations.

 If we set $\epsilon_0 = \p^s \epsilon_2$ and restrict to a
 surface $t=const$, we obtain `internal' $U(1)$ gauge
 transformations:
 \begin{equation}
 \delta \tilde{A}_{\bar{\mu}}(X,s) = \p_{\bar{\mu}} \epsilon_2(X,s)
 \end{equation}
 where $\bar{\mu}$ runs over $1,2,3,s$. These are the symmetries
 of the internal Maxwell tensor
 $\tilde{F}_{\bar{\mu} \bar{\nu}} = \p_{[ \, \bar{\mu}}
 \tilde{A}_{\bar{\nu} \, ]}$.

\subsection{Internal symmetries vs. external symmetries}

 An arbitrary history satisfies the internal equations of
 motion. Therefore all histories that are related by internal
 gauge transformations should be regarded as physically
 equivalent. However, most histories will not satisfy the
 external equations of motion, and so need not respect the
 external symmetry transformations. Histories which
 are solutions to the external equations of motion are invariant
 under both internal and external symmetry transformations.

 An internal local symmetry transformation is also an external
 local symmetry transformation if and only if
 \begin{equation}
 \label{eomsym}
 (\p^t - \p^s) \epsilon_2^s(X) = 0
 \end{equation}
 so the equation of motion $(\p^t - \p^s) \tilde{A}_\mu = 0$ is
 conserved by these transformations. It is interesting to note
 that if equation (\ref{eomsym}) holds and we set $\epsilon_0
 = \p^s \epsilon_2$ then we obtain $U(1)$ gauge
 transformations on $\N$:
 \begin{equation}
 \delta \tilde{A}_M(X,s) = \p_M \epsilon_2^s(X)
 \end{equation}
 which are the symmetries of the five-dimensional Maxwell tensor
 $\tilde{F}_{MN} = \p_{[ \, M}
 \tilde{A}_{N \, ]}$.

\section{Summary and Conclusion}

 We have shown that the global symmetry
 transformations of geometric objects in a history theory
 suggests the introduction of an extra pair of
 fields. The extended history fields can then be interpreted as fields
 on $\N = \M \times \Real$ that are covariant under the action of
 two Poincare groups, but not the $SO(2,3)$ group associated with
 \N. The history fields can be decomposed into
 two $SO(1,3)$-vectors at each point in the extended space-time,
 where the spatial components of these two vectors are the same.

\smallskip
 In the case of the massive vector field, the theory contains
 two pairs of second class constraints, and in the case of
 electromagnetism, we
 obtain three first class constraints. In both these examples,
 the extra degrees of freedom can be set to zero, and eliminated
 from the theory by taking Dirac brackets. It seems reasonable to
 expect this pattern to continue in the extension to other
 constrained field theories as there is no physical information in
 the extra degrees of freedom. So although Lorentz covariance
 suggests that the fields should have five components, the extra
 constraints allow the extra variables to be eliminated.

\smallskip
 In conclusion, the geometry of classical history theories is not fully
 understood. In the state space approach the solutions to the
 field equations are sections
 of tensor bundles associated to an $SO(1,3)$ principal bundle
 over \M. This formulation elegantly characterises the way that
 the fields transform under Lorentz transformations. If the
 history fields were covariant under $SO(2,3)$ then we would have
 a description of history fields in terms of sections of bundles
 associated to an $SO(2,3)$ principal bundle over \N. However the
 reality of the situation appears to be more complicated. There is an
 $SO(1,3)$ group associated to each surface of constant $t$, and
 to each surface of constant $s$, but the action of these two
 groups is intertwined in a non-trivial way. Similar remarks apply
 to the local symmetries in a  history theory: they cannot be
 interpreted as the transformations of a $U(1)$ connection on a
 principal bundle over \N.

\bigskip

 In a subsequent paper we will discuss the quantisation of histories
 electromagnetism using the BRST formalism.

\section{Acknowledgements}

 I would like to thank Chris Isham for his constant encouragement
 and support, and PPARC for a studentship.



\begin{thebibliography}{99}

\bibitem{Ish1}
C.J.~Isham.
\newblock Quantum Logic and the histories approach to quantum theory.
\newblock \emph{J. Math. Phys.} 35 : 2157.
\newblock 1994.
\newblock gr-qc/9308006.

\bibitem{Ish2}
C.J.~Isham and N.~Linden.
\newblock Continuous histories and the history group in generalised quantum
theory.
\newblock \emph{J. Math. Phys.} 36 : 5392.
\newblock 1995.
\newblock gr-qc/9503063.

\bibitem{Ish3}
C.J.~Isham, N.~Linden, K.~Savvidou and S.~Schreckenberg.
\newblock Continuous Time and Consistent Histories.
\newblock \emph{J. Math. Phys.} 37 : 2261.
\newblock 1998.
\newblock quant-ph/9711031.

\bibitem{Ish4}
C.J.~Isham and K.~Savvidou.
\newblock Quantising the foliation vector in Histories quantum
field theory.
\newblock In preparation

\bibitem{sav1}
K.~Savvidou
\newblock Continuous time in consistent histories (PhD thesis)
\newblock gr-qc/9912076

\bibitem{sav2}
K.~Savvidou
\newblock The action operator in continuous-time histories.
\newblock \emph{J. Math. Phys.} 40: 5657, (1999).

\bibitem{sav3}
K.~Savvidou
\newblock Poincare invariance for continuous-time histories.
\newblock gr-qc/0104053

\bibitem{sav4}
K.~Savvidou
\newblock General Relativity histories theory.
\newblock gr-qc/0104081

\bibitem{anas}
C.~Anastopoulos.
\newblock Quantum Fields in Nonstatic background: a Histories Perspective.
\newblock \emph{J. Math. Phys.} 41 : 617.
\newblock 2000.
\newblock gr-qc/9903026.

\bibitem{nolt}
D.~Noltingk
\newblock A consistent histories approach to the Unruh effect.
\newblock \emph{Int. J. Theo. Phys.} 08
\newblock gr-qc/0005063

\bibitem{chang}
S.J.~Chang.
\newblock \emph{Introduction to quantum field theory}.
\newblock Lecture Notes in Physics - Vol. 29,
\newblock World Scientific,
\newblock 1990.

\bibitem{tyutin}
D.M.~Gitman and I.V.~Tyutin.
\newblock \emph{Quantization of fields with constraints},
\newblock Berlin: Springer,
\newblock 1990.

\bibitem{henn}
M.~Henneaux and C.~Teitelboim.
\newblock \emph{Quantization of gauge systems},
\newblock Princeton University Press,
\newblock 1992.

\end{thebibliography}
\end{document}